\def\titlep{Universal fermionization of bosons
on permutative representations of the Cuntz algebra ${\cal O}_{	2}$}
\documentclass[11pt]{article}
\usepackage{graphicx,ifthen}
\usepackage{amssymb}

 at12pt

\newcommand{\sdag}{\scriptsize \dag}
\newcommand{\qed}{\hbox{\rule[-2pt]{3pt}{6pt}}}
\newcommand{\qedh}{\hfill\qed \\}

\newtheorem{Thm}{Theorem}[section]

\newtheorem{rem}[Thm]{Remark}

\newtheorem{defi}[Thm]{Definition}
\newtheorem{lem}[Thm]{Lemma}

\newtheorem{prop}[Thm]{Proposition}
\newtheorem{prob}[Thm]{Problem}

\def\cal#1{\mathcal #1}
\def\con{{\cal O}_{N}}
\def\coni{{\cal O}_{\infty}}
\def\edot{=1,\ldots,N}
\def\pr{{\it Proof.}\quad}

\def\scm#1{S({\bf C}^{N})^{\otimes #1}}

\def\co#1{{\cal O}_{#1}}
\def\disp#1{{\displaystyle #1}}
\def\brl{branching law}
\def\bfsnl{{\rm BFS}_{N}(\Lambda)}

\addtocounter{footnote}{1}
\def\sftt#1{
\setcounter{equation}{0}
\addtocounter{footnote}{1}
\section{#1}
}

\def\ssft#1{\subsection{#1}}

\begin{document}
%
% Personal data
%
\def\autherp{Katsunori Kawamura}
\def\emailp{e-mail: kawamura@kurims.kyoto-u.ac.jp.}
\def\addressp{{\it {\small College of Science and Engineering Ritsumeikan University,}}\\
{\it {\small 1-1-1 Noji Higashi, Kusatsu, Shiga 525-8577, Japan}}
}

\def\infw{\Lambda^{\frac{\infty}{2}}V}
\def\zhalfs{{\bf Z}+\frac{1}{2}}
\def\ems{\emptyset}
\def\pmvac{|{\rm vac}\!\!>\!\! _{\pm}}
\def\vac{|{\rm vac}\rangle _{+}}
\def\dvac{|{\rm vac}\rangle _{-}}
\def\ovac{|0\rangle}
\def\tovac{|\tilde{0}\rangle}
\def\expt#1{\langle #1\rangle}
\def\zph{{\bf Z}_{+/2}}
\def\zmh{{\bf Z}_{-/2}}
\def\brl{branching law}
\def\bfsnl{{\rm BFS}_{N}(\Lambda)}
\def\scm#1{S({\bf C}^{N})^{\otimes #1}}
\def\mqb{\{(M_{i},q_{i},B_{i})\}_{i=1}^{N}}
\def\zhalf{\mbox{${\bf Z}+\frac{1}{2}$}}
\def\zmha{\mbox{${\bf Z}_{\leq 0}-\frac{1}{2}$}}
\newcommand{\mline}{\noindent
\thicklines
\setlength{\unitlength}{.1mm}
\begin{picture}(1000,5)
\put(0,0){\line(1,0){1250}}
\end{picture}
\par
 }
\def\sd#1{#1^{\sdag}}
\def\dlim{{\cal D}\mbox{-}\lim}
\def\dsum{{\cal D}\mbox{-}\sum}
\def\fsum{F\mbox{-}\sum}
\def\bx{\mbox{\boldmath$x$}}
\def\wcz{\widetilde{\co{2}^{0}}}

%
%%%%%%%%% Cut from here %%%%%%%%%%
%\input comm.txt
%%%%%%%%% End of Cut %%%%%%%%%
%
%
\setcounter{section}{0}
\setcounter{footnote}{0}
\setcounter{page}{1}
\pagestyle{plain}

%%%%%%%%%%%%%%%%%%%%%%%%%%%%%%%%%%%%%%%%%%%%%%%%%%%%%%%%%%%
%
% Title
%
\title{\titlep}
\author{\autherp\thanks{\emailp}
\\ 
\addressp
}
\date{}
\maketitle

%%%%%%%%%%%%%%%%%%%%%%%%%%%%%%%%%%%%%%%%%%%%%%%%%%%%%%%%%%%
%
% Abstract
%
\begin{abstract}
Bosons and fermions are described 
by using canonical generators of Cuntz algebras
on any permutative representation.
We show a fermionization of bosons which universally holds
on any permutative representation of the Cuntz algebra ${\cal O}_{2}$.
As examples,
we show fermionizations on the Fock space and the infinite wedge.
\end{abstract}

\noindent
{\bf Mathematics Subject Classifications (2000).} 46K10, 46L60
\\
\\
{\bf Key words.} Cuntz algebra, universal fermionization

%%%%%%%%%%%%%%%%%%%%%%%%%%%%%%%%%%%%%%%%%%%%%%%%%%%%%%%%%%%%%%%%%%
%
% Section 1
%
\sftt{Introduction}
\label{section:first}
We have studied bosonization, fermionization
and boson-fermion correspondence by using 
the representation theory of operator algebras.
In this section, we explain known fermionizations of bosons
and our main theorem.
We start with problems in fermionization of bosons.
%%%%%%%%%%%%%%%%%%%%%%%%%%%%%%%%%%%%%%%%%%%%%%%%%%%%%%%%%%%%%%%%%%%%%%%%
%
% subsection 1.1
%
\ssft{What is a fermionization of bosons?}
\label{subsection:firstone}
It is often said that bosons can be written by using fermions
in various contexts \cite{Kamani,Nakawaki,Polchinski}.
Such a description is 
called a {\it fermionization of bosons} in the broad sense of the term.
This never means that the $*$-algebras of bosons 
is embedded into the $*$-algebra of fermions as a $*$-subalgebra.
Furthermore, neither a C$^{*}$-algebra 
nor a von Neumann algebra generated by fermions   
contains the $*$-algebras of bosons (\cite{RBS01}, $\S$ 1.1.1).
Such a fermionization means a description of bosons 
by using {\it not only} fermions {\it but also} the normal order with respect to
its representation \cite{MJD,Oko01}, or the description 
of Weyl forms of bosons by using a certain operator algebra of 
fermions \cite{IT}.
In consequence, any known fermionization of bosons 
{\it must} depend on some representation of the algebra of fermions
even if the representation is not apparently mentioned.

In consideration of these facts, our question is stated as follows.
%
% Problem 1.1
%
\begin{prob}
\label{prob:first}
Is there a fermionization of bosons
which does not depend on the choice of representation of 
the algebra of fermions? 
\end{prob}

\noindent
The aim of this paper is to give a weak solution for Problem \ref{prob:first}.
The new idea is that our fermionization has a kind of universality,
which is derived from the representation theory of a Cuntz algebra.

%%%%%%%%%%%%%%%%%%%%%%%%%%%%%%%%%%%%%%%%%%%%%%%%%%%%%
%
%  subsection 1.2
%
\ssft{Recursive boson system and recursive fermion system}
\label{subsection:firsttwo}
In order to define our fermionization,
we briefly explain recursive boson system and recursive fermion system 
in this subsection.
Let $\co{2}$ denote the Cuntz algebra \cite{C} with 
canonical generators $\{t_{1},t_{2}\}$, that is,
they satisfy that
%
% Equation 1.1 
% 
\begin{equation}
\label{eqn:otwo}
t_{i}^{*}t_{j}=\delta_{ij}I\quad  (i,j=1,2),\quad  t_{1}t_{1}^{*}+
t_{2}t_{2}^{*}=I. 
\end{equation}
For ${\bf N}\equiv \{1,2,3,\ldots\}$,
let $\{b_{n}:n\in {\bf N}\}$ and $\{a_{n}:n\in {\bf N}\}$
denote bosons and fermions, that is,
%
% Equation 1.3, 1.4
% 
\begin{eqnarray}
\label{eqn:boson}
\,b_{n}b_{m}^{*}-b_{m}^{*}b_{n}=\delta_{nm}I,\quad  &b_{n}b_{m}-b_{m}b_{n}
=b_{n}^{*}b_{m}^{*}-b_{m}^{*}b_{n}^{*}=0,\\
\nonumber
\\
\label{eqn:fermion}
a_{n}a_{m}^{*}+a_{m}^{*}a_{n}=\delta_{nm}I,\quad &
a_{n}a_{m}+a_{m}a_{n}=a_{n}^{*}a_{m}^{*}+a_{m}^{*}a_{n}^{*}=0
\end{eqnarray}
for each $n,m\in {\bf N}$. 
We can describe $\{b_{n}:n\in {\bf N}\}$ and $\{a_{n}:n\in {\bf N}\}$
by using $\{t_{1},t_{2}\}$ as follows \cite{BFO01}:
%
% Equation 1.4,1.5
%
\begin{eqnarray}
\label{eqn:rbszero}
b_{1}= \sum_{m\geq 1}\sqrt{m}\,t_{2}^{m-1}t_{1}t_{1}^{*}(t_{2}^{*})^{m},
\quad \,\,&b_{n}= \rho(b_{n-1})\quad(n\geq 2),\\ 
\nonumber
\\
\label{eqn:rfszero}
a_{1}= t_{1}t_{2}^{*},\,\,\quad\qquad\qquad & a_{n}=\zeta(a_{n-1})\quad(n\geq 2)
\end{eqnarray}
where 
%
% Equation 1.6, 1.7
%
\begin{eqnarray}
\label{eqn:rho}
\rho(x)= &\disp{\sum_{m\geq 1}t^{m-1}_{2}t_{1}xt_{1}^{*}(t_{2}^{*})^{m-1}},\\
\nonumber
\\
\zeta(x)= &t_{1}xt_{1}^{*}-t_{2}xt_{2}^{*}
\label{eqn:zeta}
\end{eqnarray}
for $x\in \co{2}$.
We call these descriptions of $\{b_{n}:n\in {\bf N}\}$ 
and $\{a_{n}:n\in {\bf N}\}$
as the {\it recursive boson system (=RBS)} and
the {\it recursive fermion system (=RFS)}, respectively. 

Remark that $\{b_{n}:n\in {\bf N}\}$ in (\ref{eqn:rbszero}) 
and $\rho(x)$ in (\ref{eqn:rho}) are not well-defined in $\co{2}$, 
but they make sense as operators on the reference subspace 
of any permutative representation of $\co{2}$ by Remark 1.1 in \cite{BFO01},
which are defined as follows.
%
% Definition 1.2
% 
\begin{defi}
\label{defi:first}
\cite{BJ,DaPi2,DaPi3,RBS01}
A representation $({\cal H},\pi)$ of $\co{2}$
is permutative if 
there exists an orthonormal basis ${\cal E}\equiv \{e_{n}:n\in \Lambda\}$ 
of ${\cal H}$
such that $\pi(t_{i}){\cal E}\subset {\cal E}$ for each $i=1,2$.
We call ${\cal E}$
and the linear hull ${\cal D}$ of ${\cal E}$
by the reference basis and the reference subspace 
of $({\cal H},\pi)$, respectively.
\end{defi}

\noindent
Any permutative representation is decomposed into 
the direct sum of cyclic permutative representations
unique up to unitary equivalence.
Unitary equivalence classes of irreducible permutative representations 
are completely classified.
There exist uncountably many unitary equivalence classes 
of irreducible permutative representations \cite{BJ,DaPi2,DaPi3}.

The RFS and the RBS induce branching laws of $*$-representations 
of Cuntz algebras on fermions and bosons \cite{AK1,AK05,AK02R,IWF01,RBS01,BFO01},
and a nonlinear transformation group of fermions \cite{AK06}.

%%%%%%%%%%%%%%%%%%%%%%%%%%%%%%%%%%%%%%%%%%%%%%%%%%%
%
% subsection 1.3
%
\ssft{Main theorem}
\label{subsection:firstthree}
In this subsection, we show our main theorem.
Let $t_{1},t_{2}$ denote canonical generators of $\co{2}$
in (\ref{eqn:otwo}).
Assume that they act on a permutative representation 
with the reference subspace ${\cal D}$.
Let ${\cal A}$ denote the $*$-algebra generated by fermions 
$\{a_{n}:n\in {\bf N}\}$
in (\ref{eqn:rfszero}).
Define the subset $\{W_{n}:n\geq 0\}$ of ${\cal A}$ by
%
% Equation 1.8
%
\begin{equation}
\label{eqn:woperator}
W_{0}\equiv a_{1}a_{1}^{*},\quad
W_{n}\equiv a_{n+1}a_{n+1}^{*}a_{n}^{*}a_{n}\cdots a_{1}^{*}a_{1}\quad(n\geq 1).
\end{equation}
Then we can verify that $\{W_{n}:n\geq 0\}$ 
is an orthogonal family of projections, that is,
%
% Equation 1.9
%
\begin{equation}
\label{eqn:projections}
W_{n}^{*}=W_{n},\quad
W_{n}^{2}=W_{n},\quad W_{n}W_{m}=0\quad(n\ne m),
\end{equation}
on ${\cal D}$ such that $\sum_{n\geq 0}W_{n}=I$ on ${\cal D}$.
%
% Lemma 1.3
%
\begin{lem}
\label{lem:finite}
\begin{enumerate}
%(i)
\item
For $\rho$ in (\ref{eqn:rho}),
$\rho(a_{n})=\sum_{m\geq 0}(-1)^{m}a_{n+m+1}W_{m}$
for $n\geq 1$
where the formal infinite sum is actually finite on ${\cal D}$.
%(ii)
\item
For any operators $x,y$ on ${\cal D}$,
$\rho(xy)=\rho(x)\rho(y)$ on ${\cal D}$.
%(iii)
\item
Define the formal infinite sum $Y$ of fermions by
%
% Equation 1.10
%
\begin{equation}
\label{eqn:yoperator}
Y\equiv a_{1}^{*}a_{2}+\sum_{n\geq 2}a_{1}^{*}a_{1}\cdots a_{n-1}^{*}a_{n-1}a_{n}^{*}a_{n+1}.
\end{equation}
Then the sum in (\ref{eqn:yoperator}) is actually finite on ${\cal D}$
and $Y{\cal D}\subset {\cal D}$.
\end{enumerate}
\end{lem}
From Lemma \ref{lem:finite}(i),
if $x$ is a formal infinite sum of fermions acting on ${\cal D}$,
then so is $\rho(x)$. 
%
% Theorem 1.4
%
\begin{Thm}
\label{Thm:main}
Let ${\cal D}$ denote the reference subspace of a permutative
representation of $\co{2}$ and let $t_{1},t_{2}$ denote
canonical generators of $\co{2}$ acting on ${\cal D}$.
Let $\rho$, $W_{n}$, $Y$ be as in 
(\ref{eqn:rho}),  (\ref{eqn:projections}) and (\ref{eqn:yoperator}), respectively.
For $n\geq 1$,
define the set $\{F_{n}:n\geq 1\}$ of formal infinite sums of fermions by
%
% Equation 1.11
%
\begin{equation}
\label{eqn:cluster}
F_{1}\equiv 
\sum_{m\geq 1}\sqrt{m}\,W_{m},\quad
F_{n}\equiv Y\rho(F_{n-1})\quad(n\geq 2).
\end{equation}
Then the following holds:
\begin{enumerate}
%(i)
\item
Formal infinite sums in $F_{n}$'s are actually finite on ${\cal D}$
and $F_{n}{\cal D}\subset {\cal D}$ for each $n\geq 1$.
%(i)
\item
For $n\geq 1$, a boson $b_{n}$ in (\ref{eqn:rbszero})
is written as the product of $t_{2}^{*}$ and $F_{n}$:
%
% Equation 1.12
%
\begin{equation}
\label{eqn:maineq}
b_{n}=t_{2}^{*}F_{n}.
\end{equation}
\end{enumerate}
\end{Thm}

\noindent
Theorem \ref{Thm:main} shows that 
bosons are written by using fermions and one of (the $*$-conjugate of)
canonical generators, $t_{2}^{*}$ of $\co{2}$.
The universality of the fermionization of bosons (\ref{eqn:maineq}) holds
on the reference subspace of {\it any} permutative representation of $\co{2}$.
Hence a weak solution of Problem \ref{prob:first} is given.
We see that 
a boson $b_{n}$ corresponds with the ``cluster" $F_{n}$ of fermions
by the canonical generator $t_{2}^{*}$ of $\co{2}$ in
a purely algebraic sense.
%
% Remark 1.5
%
\begin{rem}
\label{rem:one}
{\rm
\begin{enumerate}
%(i)
\item
The formula (\ref{eqn:maineq}) is derived from
the RBS (\ref{eqn:rbszero}), the RFS (\ref{eqn:rfszero}) 
and permutative representations of $\co{2}$
but not from a specific physical model.
Especially,
our fermionization has no relation with the dimension of any space-time.
%(ii)
\item
Instead of normal product or operator topologies,
we use $t_{2}^{*}$ in (\ref{eqn:maineq}).
Hence $b_{n}$ is not written by using only fermions.
%(iii)
\item
The fermionization (\ref{eqn:maineq}) holds 
on any permutative representation
even if it is not irreducible.
%(iv)
\item
Theorem \ref{Thm:main} shows that 
any boson is written as (the limit of)
even numbers of fermions except $t_{2}^{*}$.
\end{enumerate}
}
\end{rem}

According to \cite{BFO01},
we show the significance of the
fermionization (\ref{eqn:maineq}).
For a given permutative representation $\pi$ of $\co{2}$,
we obtain the restriction $\pi|_{{\cal A}}$ of $\pi$
on the $*$-algebra ${\cal A}$ of fermions.
From the fermionization (\ref{eqn:maineq}),
we obtain the representation $(\pi|_{{\cal A}})|_{{\cal B}}$ of 
bosons. The operation
%
% Equation 1.13
%
\begin{equation}
\label{eqn:restriction}
\pi|_{{\cal A}}\mapsto (\pi|_{{\cal A}})|_{{\cal B}}
\end{equation}
seems that $(\pi|_{{\cal A}})|_{{\cal B}}$ 
was the restriction of $\pi|_{{\cal A}}$ on ${\cal B}$
which is the $*$-algebra of bosons.
Remark that $\pi|_{{\cal A}}$
is not irreducible even if $\pi$ is irreducible.
The operation (\ref{eqn:restriction}) holds
on any permutative representation $\pi$ of $\co{2}$,
but does not always hold on representation of fermions.

At the last of this section,
we consider the inverse formula of Theorem \ref{Thm:main}.
When fermions are written by using bosons,
such a description is 
called a {\it bosonization of fermions} in the broad sense of the term
\cite{AMS,Kamani,Kopietz,Stone}.
Theorem \ref{Thm:main} brings the following natural question. 
%
% Problem 1.6
%
\begin{prob}
\label{prob:second}
In analogy with Theorem \ref{Thm:main},
find bosonization formulae of fermions,
which is universally holds
on the reference subspace of {\it any} permutative representation of $\co{2}$.
\end{prob}

In $\S$ \ref{section:second}, we prove Lemma \ref{lem:finite} 
and Theorem \ref{Thm:main}.
In $\S$ \ref{section:third}, we show examples of Theorem \ref{Thm:main}.

%%%%%%%%%%%%%%%%%%%%%%%%%%%%%%%%%%%%%%%%%%%%%%%%%%%%%%%%%%%%%%%%
%
% Section 2
%
\sftt{Proofs of theorems}
\label{section:second}
In this section, we prove Lemma \ref{lem:finite} and Theorem \ref{Thm:main}.
%%%%%%%%%%%%%%%%%%%%%%%%%%%%%%%%%%%%%%%%%%%%%%%%%%%%%%%%%%%%%%%
%
%  subsection 2.1
%
\ssft{Permutative representations of Cuntz algebras}
\label{subsection:secondone}
For $N=2,3,\ldots,+\infty$, 
let $\con$ denote the {\it Cuntz algebra} \cite{C}, that is, a C$^{*}$-algebra 
which is universally generated by $s_{1},\ldots,s_{N}$ satisfying
$s_{i}^{*}s_{j}=\delta_{ij}I$ for $i,j\edot$ and
\[\sum_{i=1}^{N}s_{i}s_{i}^{*}=I\quad(\mbox{if } N<+\infty),\quad
\sum_{i=1}^{k}s_{i}s_{i}^{*}\leq I,\quad k= 1,2,\ldots\quad
(\mbox{if }N = +\infty)\]
where $I$ denotes the unit of $\con$.

Let $\{t_{1},t_{2}\}$ and $\{s_{n}:n\in {\bf N}\}$
denote canonical generators of $\co{2}$ and $\coni$, respectively.
We introduce two (classes of) permutative representations of $\co{2}$.
%
% Definition 2.1
% 
\begin{defi}
\label{defi:permudef}
\cite{AK1,AK02R,IWF01,BFO01}
We write a class $P(1)$ ({\it resp}. $P(12)$)
of representations $({\cal H},\pi)$ of $\co{2}$ with  
a cyclic unit vector $\Omega\in {\cal H}$ such that $\pi(t_{1})\Omega=\Omega$
({\it resp}. $\pi(t_{1}t_{2})\Omega=\Omega$).
%We call $\Omega$ the GP vector of $({\cal H},\pi)$.
\end{defi}

\noindent
Both $P(1)$ and $P(12)$ contain only one unitary equivalence class.
Hence we identify $P(1)$ and $P(12)$ with their representatives,
respectively.
They are irreducible and permutative but not unitarily equivalent.

By using the embedding of $\coni$ into $\co{2}$ defined by
%
% Equation 2.1
%
\begin{equation}
\label{eqn:embedding}
s_{n}=t_{2}^{n-1}t_{1}\quad(n\geq 1),
\end{equation}
the restriction 
of any permutative representation of $\co{2}$ on $\coni$
is also permutative.
Furthermore, if ${\cal D}$ is the reference subspace
of a permutative representation of $\co{2}$,
then ${\cal D}$ is that of $\coni$.
Then the following holds.
%
% Lemma 2.2
%
\begin{lem}
\label{lem:vanish}
Let $\{s_{n}:n\in {\bf N}\}$ be as in (\ref{eqn:embedding}) and
let ${\cal D}$ denote the reference subspace of a permutative representation
of $\co{2}$.
Then, for any $v\in {\cal D}$,
$s_{n}^{*}v=0$
except finite number of $n\in {\bf N}$.
\end{lem}
%
% Proof
%
\pr
By definitions of permutative representation
and the reference subspace,
we see that
${\cal D}=\bigcup_{n\geq 1} s_{n}{\cal D}$
and $s_{n}{\cal D}\cap s_{m}{\cal D}=\{0\}$ when $n\ne m$.
For any $v\in {\cal D}$, there exists a nonempty finite subset $A$ of ${\bf N}$
such that $v\in \oplus_{k\in A}s_{k}{\cal D}$.
Hence $s_{n}^{*}v=0$ when $n\not\in A$. Therefore the statement holds.
\qedh

\noindent
The essential part of our fermionization is 
derived from (\ref{eqn:embedding}).
The relation (\ref{eqn:embedding}) 
also appears in the metric theory of continued fractions \cite{CFR01}.

%%%%%%%%%%%%%%%%%%%%%%%%%%%%%%%%%%%%%%%%%%%%%%%%%
%
% subsection 2.2
%
\ssft{Proofs of Lemma \ref{lem:finite} and Theorem \ref{Thm:main}}
\label{subsection:secondtwo}
Define the subset $\{X_{n}:n\geq 1\}$ of ${\cal A}$ by
%
% Equation 2.2
%
\begin{equation}
\label{eqn:xoperator}
X_{1}\equiv a_{1}^{*}a_{2}, \quad
X_{n}\equiv a_{1}^{*}a_{1}\cdots a_{n-1}^{*}a_{n-1}a_{n}^{*}a_{n+1}
\quad (n\geq 2).
\end{equation}
By definition, the following holds.
%
% Lemma 2.3
%
\begin{lem}
\label{lem:formula}
Let $\rho$, $W_{n}$, $s_{n}$, $X_{n}$ be as in
(\ref{eqn:rho}), (\ref{eqn:woperator}), (\ref{eqn:embedding}) 
and (\ref{eqn:xoperator}), respectively.
Then the following holds:
\begin{enumerate}
%(i)
\item
For each $n\geq 1$, $t_{2}s_{n}=s_{n+1}$.
%(ii)
\item
For each $n\geq 0$, $W_{n}=s_{n+1}s_{n+1}^{*}$.
%(iii)
\item
For each $n\geq 1$, $s_{n}t_{2}^{*}s_{n}^{*}=t_{2}^{*}X_{n}$.
%(iv)
\item
For  $\{a_{n}\}$ in (\ref{eqn:rfszero}),
we obtain the following:
\[s_{m}a_{n}=(-1)^{m-1}a_{n+m}s_{m},
\quad
s_{m}a_{n}^{*}=(-1)^{m-1}a_{n+m}^{*}s_{m}
\quad(n,m\geq 1).\]
\end{enumerate}
\end{lem}

\noindent
{\it Proof of Lemma \ref{lem:finite}.}
(i)
By definition,
$\rho(a_{n})=\sum_{m\geq 1}s_{m}a_{n}s_{m}^{*}$.
From Lemma \ref{lem:formula}(ii) and (iv),
$s_{m}a_{n}s_{m}^{*}=(-1)^{m-1}a_{n+m}W_{m-1}$
for $m\geq 1$.
For any $v\in {\cal D}$,
$W_{n}v=0$ except finite number of $n\in {\bf N}$
from Lemma \ref{lem:vanish} and  Lemma \ref{lem:formula}(ii).
Hence 
the sum $\sum_{m\geq 1}(-1)^{m-1}a_{n+m}W_{m-1}$ 
is actually finite on ${\cal D}$.

\noindent
(ii)
Let $v\in {\cal D}$.
From Lemma \ref{lem:vanish},
there exists a nonempty finite subset $A$ of ${\bf N}$ such that
$s_{n}^{*}v=0$ when $n\not\in A$.
Hence 
%
% Equation 2.3
%
\begin{equation}
\label{eqn:rhoone}
\rho(xy)v
=\sum_{m\geq 1}s_{m}xys_{m}^{*}v
=\sum_{m\in A}s_{m}xys_{m}^{*}v.
\end{equation}
On the other hand,
$\rho(y)v=\sum_{m\in A}s_{m}ys_{m}^{*}v\in {\cal D}$.
From this,
%
% Equation 2.4
%
\begin{equation}
\label{eqn:rhotwo}
\rho(x)\rho(y)v
=\rho(x)\sum_{m\in A}s_{m}ys_{m}^{*}v
=\sum_{m\in A}\rho(x)s_{m}ys_{m}^{*}v
=\sum_{m\in A}s_{m}xys_{m}^{*}v.
\end{equation}
Hence the statement holds.

\noindent
(iii)
By definition and Lemma \ref{lem:formula}(iii),
$Y=\sum_{n\geq 1}X_{n}=\sum_{n\geq 1}s_{n+1}t_{2}^{*}s_{n}^{*}$.
From Lemma \ref{lem:vanish}, the sum in $Y$ is actually finite on ${\cal D}$.
Since $s_{n+1}t_{2}^{*}s_{n}{\cal D}\subset {\cal D}$,
$Y{\cal D}\subset {\cal D}$.
\qedh

\noindent
{\it Proof of Theorem \ref{Thm:main}.}
(i)
From Lemma \ref{lem:formula}(ii) and Lemma \ref{lem:vanish},
the sum in $F_{1}$ is actually finite on ${\cal D}$.
From Lemma \ref{lem:finite}(iii) and Lemma \ref{lem:vanish},
the sum in $\rho(F_{1})$ is actually finite on ${\cal D}$.
From this, $F_{2}=Y\rho(F_{1})$ is actually finite on ${\cal D}$.
In this way, we see that the sum in $F_{n}$
is actually finite on ${\cal D}$ for all $n\geq 1$.

\noindent
(ii)
From Lemma \ref{lem:formula}(iii) and Lemma \ref{lem:finite}(iii),
we see that
%
% Equation 2.5
%
\begin{equation}
\label{eqn:rhoy}
\rho(t_{2}^{*})
=\sum_{n\geq 1}s_{n}t_{2}^{*}s_{n}^{*}=t_{2}^{*}\sum_{n\geq 1}X_{n}=t_{2}^{*}Y
\end{equation}
for $Y$ in (\ref{eqn:yoperator}).

From (\ref{eqn:rbszero}) and Lemma \ref{lem:formula}(ii),
we obtain that 
$b_{1}=t_{2}^{*}\sum_{n\geq 1}\sqrt{n}\,W_{n}$
where the formal infinite sum of fermions is actually finite on ${\cal D}$
from Lemma \ref{lem:vanish}.
Hence the case $n=1$ holds.
Assume that (\ref{eqn:maineq}) holds for $n\leq n_{0}$
for some $n_{0}\in {\bf N}$.
From Lemma \ref{lem:finite}(ii) and (\ref{eqn:rbszero}),
%
% Equation 2.5
%
\begin{equation}
\label{eqn:bformula}
b_{n_{0}+1}=\rho(b_{n_{0}})
=\rho(t_{2}^{*}F_{n_{0}})
=\rho(t_{2}^{*})\rho(F_{n_{0}}).
\end{equation}
From (\ref{eqn:rhoy}) and the definition of $F_{n}$,
we obtain that
$b_{n_{0}+1}=t_{2}^{*}F_{n_{0}+1}$.
By the inductive method, the statement holds.
\qedh
%%%%%%%%%%%%%%%%%%%%%%%%%%%%%%%%%%%%%%%%%%%%%%%%%%%%%%%%%%%%%%
%
% Section 3
%
\sftt{Examples}
\label{section:third}
We show examples of Theorem \ref{Thm:main} in this section.
%%%%%%%%%%%%%%%%%%%%%%%%%%%%%%%%%%%%%%%%
%
% subsection 3.1
%
\ssft{$F_{1}$ and $F_{2}$}
\label{subsection:thirdone}
We show more concrete representations of $F_{1}$ and $F_{2}$
in (\ref{eqn:cluster}) as follows.
%
% Proposition 3.1
%
\begin{prop}
\label{prop:fone}
%
% Equation 3.1, 3.2
%
\begin{eqnarray}
\label{eqn:fone}
F_{1}=&\disp{\sum_{n\geq 1}\sqrt{n}
a_{1}^{*}a_{1}\cdots a_{n}^{*}a_{n}a_{n+1}a_{n+1}^{*},}\\
\nonumber
\\
\nonumber
F_{2}=&\disp{\sum_{n,m\geq 1}\sqrt{m}a_{1}^{*}a_{1}\cdots a_{n-1}^{*}a_{n-1}
a_{n}^{*}a_{n+1}a_{n+2}^{*}a_{n+2}\cdots a_{n+m}^{*}a_{n+m}a_{n+m+1}a_{n+m+1}^{*}}\\
\label{eqn:ftwo}
\end{eqnarray}
where we write $a_{0}=a_{0}^{*}=I$.
\end{prop}
%
% Proof
%
\pr
By definition, (\ref{eqn:fone}) holds.
For $n\leq m$,
define $R[n,m]\equiv a_{n}^{*}a_{n}\cdots a_{m}^{*}a_{m}$.
Then we can verify that
$\rho(W_{m})=\sum_{l\geq 0}a_{m+l+2}a_{m+l+2}^{*}R[l+2,l+1+m]W_{l}$
for each $m\geq 1$.
From this,  (\ref{eqn:ftwo}) is verified.
\qedh

%%%%%%%%%%%%%%%%%%%%%%%%%%%%%%%%%%%%%%%%%%%%%%%%%%%%%%%
%
% subsection 3.2
%
\ssft{Fermionization on the Fock representation}
\label{subsection:thirdtwo}
In this subsection, we show that
the fermionization (\ref{eqn:maineq}) induces the Bose-Fock representation
from the Fermi-Fock representation.
%
% Proposition 3.2
%
\begin{prop}
\label{prop:fock}
Let ${\cal B}$ and ${\cal A}$ denote $*$-algebras generated by
$\{b_{n}:n\geq 1\}$ in (\ref{eqn:boson}) and
$\{a_{n}:n\geq 1\}$ in (\ref{eqn:fermion}), respectively.
We assume that 
${\cal A}$ is embedded into $\co{2}$ by (\ref{eqn:rfszero}).
\begin{enumerate}
%(i)
\item
Let $({\cal H}_{F},\pi_{F})$ be the Fermi-Fock representation of ${\cal A}$
with the vacuum $\Omega$,
that is, $\Omega$ is a cyclic unit vector of ${\cal H}_{F}$ such that 
%
% Equation 3.3
%
\begin{equation}
\label{eqn:fock}
\pi_{F}(a_{n})\Omega=0\quad(n\geq 1).
\end{equation}
Then there exists a cyclic action $\tilde{\pi}_{F}$ of $\co{2}$ on ${\cal H}_{F}$
such that $\tilde{\pi}_{F}|_{{\cal A}}=\pi_{F}$ and $\tilde{\pi}_{F}(t_{1})\Omega
=\Omega$.
%(ii)
\item
The cyclic representation $({\cal H}_{F},\pi_{F},\Omega)$ of ${\cal A}$
in (i) induces the Bose-Fock representation with $\Omega$
as the Bose-Fock vacuum,
with respect to the fermionization (\ref{eqn:maineq}), that is,
%
% Equation 3.4
%
\begin{equation}
\label{eqn:bfock}
\tilde{\pi}_{F}(b_{n})\Omega=0\quad(n\geq 1)
\end{equation}
and $\tilde{\pi}_{F}({\cal B})\Omega$ 
is a dense subspace of ${\cal H}_{F}$
with respect to (\ref{eqn:rbszero}).
\end{enumerate}
\end{prop}
%
% Proof
%
\pr
From Theorem 1.2(ii) of \cite{BFO01}, the former statement of (i) holds.
From (i), $({\cal H}_{F},\tilde{\pi}_{F})$ is $P(1)$ of $\co{2}$
in Definition \ref{defi:permudef}.
Hence it is permutative.
From Theorem 1.2(i) of \cite{BFO01}, the latter statement of (i) holds.
From the uniqueness of the action of $\co{2}$ in (i),
The statement (ii) holds from Theorem 1.2(iii) of \cite{BFO01}.
\qedh

\noindent
In consequence, 
the fermionization (\ref{eqn:maineq}) preserves Fock vacua of fermions and bosons.

In fact, we show (\ref{eqn:bfock}) for $n=1,2$ here.
We write $\pi_{F}(x)$, $\tilde{\pi}_{F}(y)$
as $x,y$ for the simplicity of description.
From (\ref{eqn:maineq}),
%
% Equation 3.5
%
\begin{equation}
\label{eqn:maineqtwo}
b_{n}^{*}=F_{n}^{*}t_{2}\quad(n\geq 1).
\end{equation}
Let $\Omega$ be as in (\ref{eqn:fock}).
For $n\geq 1$,
let $R_{n}\equiv a_{1}^{*}a_{1}\cdots a_{n}^{*}a_{n}$.
From (\ref{eqn:maineqtwo}),
\[
\begin{array}{rl}
b_{1}^{*}\Omega
=&F_{1}^{*}t_{2}\Omega\\
=&\sum_{n\geq 1}\sqrt{n}W_{n}t_{2}\Omega\\
=&\sum_{n\geq 1}\sqrt{n}a_{n+1}a_{n+1}^{*}R_{n}t_{2}\Omega\\
=&\sum_{n\geq 1}\sqrt{n}a_{n+1}a_{n+1}^{*}a_{n}^{*}a_{n}\cdots 
a_{2}^{*}a_{2}a_{1}^{*}\Omega\\
=&a_{2}a_{2}^{*}a_{1}^{*}\Omega\\
=&a_{1}^{*}\Omega.
\end{array}
\]
From (\ref{eqn:ftwo}),
\[
\begin{array}{rl}
b_{2}^{*}\Omega
=&F_{2}^{*}t_{2}\Omega\\
=&
\sum_{n,m\geq 1}\sqrt{m}R_{n-1}
a_{n+1}^{*}a_{n}a_{n+2}^{*}a_{n+2}\cdots a_{n+m}^{*}a_{n+m}a_{n+m+1}a_{n+m+1}^{*}
t_{2}\Omega\\
=&
\sum_{m\geq 1}\sqrt{m}
a_{2}^{*}a_{1}a_{3}^{*}a_{3}\cdots a_{1+m}^{*}a_{1+m}a_{2+m}a_{2+m}^{*}
t_{2}\Omega\\
=&
\sum_{m\geq 1}\sqrt{m}
a_{3}^{*}a_{3}\cdots a_{1+m}^{*}a_{1+m}a_{2+m}a_{2+m}^{*}a_{2}^{*}\Omega\\
=&a_{3}a_{3}^{*}a_{2}^{*}\Omega\\
=&a_{2}^{*}\Omega.
\end{array}
\]

%%%%%%%%%%%%%%%%%%%%%%%%%%%%%%%%%%%%%%%%%%%%%%%%%%%%%%%%%%%%%
%
% subsection 3.3
%
\ssft{Fermionization on the infinite wedge}
\label{subsection:thirdthree}
According to \cite{IWF01}, we show the fermionization 
on the infinite wedge.
Define ${\bf Z}_{\geq 0}+1/2\equiv \{n+1/2:n=0,1,2,\ldots\}$.
For $k\in {\bf Z}_{\geq 0}+1/2$,
rewrite $\{a_{n}:n\in {\bf N}\}$ in (\ref{eqn:fermion}) as
%
% Equation 3.6
%
\begin{equation}
\label{eqn:psi}
\psi_{k}\equiv a_{2k+1},\quad
\psi_{-k}\equiv a_{2k}.
\end{equation}
Of course,
the $*$-algebra generated by
$\{\psi_{k},\,\psi_{-k}:k\in {\bf Z}_{\geq 0}+1/2\}$ coincides with ${\cal A}$. 
Let $\Omega$ and $\Omega^{*}$ be vectors in a representation space of ${\cal A}$ 
such that
%
% Equation 3.7
%
\begin{equation}
\label{eqn:psiomega}
\psi_{-k}\Omega=\psi_{k}^{*}\Omega=0\quad(k\in {\bf Z}_{\geq 0}+1/2),
\end{equation}
%
% Equation 3.8
%
\begin{equation}
\label{eqn:psiomegatwo}
\psi_{k}\Omega^{*}=\psi_{-k}^{*}\Omega^{*}=0\quad (k\in {\bf Z}_{\geq 0}+1/2).
\end{equation}
The cyclic subrepresentation 
of ${\cal A}$ generated by $\Omega$
({\it resp}. $\Omega^{*}$) is called the {\it infinite wedge representation}
({\it resp. dual infinite wedge representation}).
From (\ref{eqn:psi}),
%
% Equation 3.9
%
\begin{equation}
\label{eqn:newone}
a_{2n-1}\Omega=a^{*}_{2n}\Omega=0\quad(n\in {\bf N}),
\end{equation}
%
% Equation 3.10
%
\begin{equation}
\label{eqn:newtwo}
a_{2n}\Omega^{*}=a_{2n-1}\Omega^{*}=0\quad(n\in {\bf N}).
\end{equation}
From \cite{IWF01}, they are irreducible
and they are not unitarily equivalent.

%
% Proposition 3.3
%
\begin{prop}
\label{prop:threethree}
\begin{enumerate}
%(i)
\item
Let $({\cal H}_{IW},\pi_{IW})$ and 
$({\cal H}_{IW^{*}},\pi_{IW^{*}})$ be
the infinite wedge representation with the vacuum $\Omega$
and the dual infinite wedge representation with the dual vacuum $\Omega^{*}$
of ${\cal A}$, respectively.
Then there exists a cyclic action $\pi_{2}$ of $\co{2}$
on ${\cal H}_{IW}\oplus {\cal H}_{IW^{*}}$
such that 
$\pi_{2}|_{{\cal A}}=\pi_{IW}\oplus \pi_{IW^{*}}$
and 
$\pi_{2}(t_{2})\Omega=\Omega^{*}$ and $\pi_{2}(t_{1})\Omega^{*}=\Omega$.
%(ii)
\item
The direct sum representation $({\cal H}_{IW}\oplus {\cal H}_{IW^{*}},
\pi_{IW}\oplus \pi_{IW^{*}})$ of ${\cal A}$
induces the following representation 
$\eta\equiv (\pi_{IW}\oplus \pi_{IW}^{*})|_{{\cal B}}$  of bosons 
with the cyclic vector $\Omega$ in (i),
with respect to the fermionization (\ref{eqn:maineq}):
%
% Equation 3.11
%
\begin{equation}
\label{eqn:newthree}
\eta(b_{n}b_{n}^{*})\Omega=2\Omega\quad(n\in {\bf N}).
\end{equation}
\end{enumerate}
\end{prop}
%
% Proof
%
\pr
(i)
See Theorem 1.1(ii) in \cite{IWF01}.

\noindent
(ii) 
We see that 
the cyclic representation $\eta$ in (\ref{eqn:newthree}) is $BF_{1,1}(2)$
in Definition 3.4 of \cite{BFO01}.
Furthermore we see that $\pi_{IW}$ and
$\pi_{IW^{*}}$
are $FF_{2,1}$ and $FF_{2,2}$ in Example 3.7(ii) 
of \cite{BFO01}, respectively.
From Example 4.7(ii) in \cite{BFO01},
we see that
they are the restrictions of $P(12)$ in Definition \ref{defi:permudef}.
Hence the statement holds.
\qedh
%
% Remark 3.4
%
\begin{rem}
\label{rem:infinite}
{\rm
From Proposition \ref{prop:threethree},
two irreducible representations of fermions
induce one irreducible representation of bosons
with respect to the fermionization (\ref{eqn:maineq}).
On the other hand,
the infinite wedge representation is decomposed into 
the direct sum of countably infinite many 
Bose-Fock representations 
with respect to the fermionization in \cite{MJD,Oko01}.
Hence our fermionization is different from that in \cite{MJD,Oko01}.
}
\end{rem}

%\cls
%\section*{Appendix}
%\appendix
%\input app01.txt 

%%%%%%%%%%%%%%%%%%%%%%%%%%

\end{document}